\documentclass[11pt,a4paper,cite]{article}
\usepackage[english]{babel}
\usepackage[T1]{fontenc}
\usepackage[latin1]{inputenc}
\usepackage{epsfig}
\usepackage{graphics}
\usepackage{cite}
\usepackage{amsmath}
\usepackage{amsfonts}
\usepackage{amssymb}
\usepackage{amsthm}
\usepackage
[makeroom]{cancel}
\usepackage{color}
\usepackage{bm}
\usepackage{datetime}
\usepackage{xcolor}

\newcommand{\beq}{\begin{equation}}
\newcommand{\eeq}{\end{equation}}

\newcommand\blfootnote[1]{%
	\begingroup
	\renewcommand\thefootnote{}\footnote{#1}%
	\addtocounter{footnote}{-1}%
	\endgroup
}

\newdateformat{monthyeardate}{%
	\monthname[\THEMONTH], \THEYEAR}

\title{\bf On the inexistence of solitons in Einstein-Maxwell-scalar models}
\author{Carlos A. R. Herdeiro$^{1\ddagger}$, Jo\~{a}o M. S. Oliveira$^{2\dagger}$,}
\date{%
	$^1${\small Centro de Astrof\'\i sica e Gravita\c c\~ao - CENTRA,} \\ {\small Departamento de F\'\i sica,
Instituto Superior T\'ecnico - IST, Universidade de Lisboa,} \\  {\small Avenida
Rovisco Pais 1, 1049-001, Portugal}\\ \vspace{0.3cm}
$^2${\small Departamento de F\'isica da Universidade de Aveiro and CIDMA,} \\ {\small Campus de Santiago, 3810-183 Aveiro, Portugal}\\
 \vspace{0.3cm}
	\monthyeardate\today
}

\begin{document}
\maketitle
\blfootnote{$\ddagger$ carlosherdeiro@tecnico.ulisboa.pt}
\blfootnote{$\dagger$ jmiguel.oliveira@ua.pt}

\begin{abstract}
 Three non-existence results are established for self-gravitating solitons in Einstein-Maxwell-scalar models, wherein the scalar field is, generically, non-minimally coupled to the Maxwell field via a scalar function $f(\Phi)$. Firstly,  a trivial Maxwell field is considered, which yields a consistent truncation of the full model. In this case, using a scaling (Derrick-type) argument, it is established that no stationary and axisymmetric self-gravitating scalar solitons exist, unless the scalar potential energy is somewhere negative in spacetime. This generalises previous results for the static and strictly stationary cases. Thus, rotation alone cannot support self-gravitating scalar solitons in this class of models. Secondly, constant sign couplings are considered. Generalising a previous argument by Heusler for electro-vacuum, it is established that no static self-gravitating electromagnetic-scalar solitons exist. Thus, a varying (but constant sign) electric permittivity alone cannot support static Einstein-Maxwell-scalar solitons. Finally, the second result is generalised for strictly stationary, but not necessarily static, spacetimes, using a Lichnerowicz-type argument, generalising previous results in models where the scalar and Maxwell fields are not directly coupled. The scope of validity of each of these results points out the possible paths to circumvent them, in order to obtain self-gravitating solitons in Einstein-Maxwell-scalar models. 
\end{abstract}

\newpage



\section{Introduction}
Vacuum general relativity is described by the Einstein-Hilbert action,
\beq
\mathcal{S}_{EH}=\frac{1}{2}\int d^4x \sqrt{-g} R \ ,
\eeq
where units are chosen to set Newton's constant (times $8\pi$) to unity and $R$ is the Ricci curvature scalar. One may ask whether this non-linear theory of gravity allows for equilibrium, localised, finite energy solutions. It is well known that it does; there are \textit{black hole} solutions. According to a set of uniqueness theorems~\cite{Israel:1967wq,Carter:1971zc,Robinson:1975bv} (see also~\cite{HeuslerBook,Heusler:1998ua,Chrusciel:2012jk}), the most general, non-singular on and outside an event horizon, stationary black hole is provided by the Kerr two-parameter family of metrics~\cite{Kerr:1963ud}. These metrics, however, have curvature singularities within the event horizon. Moreover,  fully regular black holes ($i.e.$ also inside the event horizon) are impossible in vacuum gravity by the same aforementioned uniqueness theorems. Then, one may ask 
 whether Einstein's vacuum general relativity allows for \textit{everywhere} non-singular, asymptotically flat, equilibrium solutions describing energy lumps without an event horizon. These would be particle-like solutions, also known as  self-gravitating \textit{solitons}. Physically, one could imagine such speculative solutions as a bundle of gravitational waves tied up under their own ``weight". Classical results of general relativity, however, rule out non-zero mass solitonic solutions admitting an asymptotically timelike Killing vector field, even for non-trivial topologies~\cite{Einstein:1943ixi,Lichnerowicz} (see also~\cite{Heusler:1996ft}). The positive mass theorem then establishes that the only zero mass solution is Minkowski spacetime~\cite{Schon:1979rg,Witten:1981mf}.

Adding to vacuum general relativity the Maxwell field $F_{\mu\nu}$, without sources, one is led to consider the (source-free) Einstein-Maxwell theory, also referred to as \textit{electro-vacuum}, which is described by the action
\beq
\mathcal{S}_{EM}=\mathcal{S}_{EH}-\frac{1}{4}\int d^4x \sqrt{-g} F_{\mu\nu}F^{\mu\nu} \ ,
\label{actionem}
\eeq
where units are chosen that set the vacuum permittivity (times $4\pi$) to unity. One may repeat the very same questions in this model, concerning the existence of localised, equilibrium, finite energy solutions. Again, black hole uniqueness theorems~\cite{Israel:1967za,Robinson:1974nf,HeuslerBook,Heusler:1998ua,Chrusciel:2012jk} establish the most general non-singular on and outside an event horizon \textit{single} black hole solutions is provided by the four parameter Kerr-Newman family~\cite{Newman:1965my}, which again rules out fully regular black holes. A new feature in electro-vacuum, however, is the existence of a class of multi-black hole solutions, the Majumdar-Papapetrou family~\cite{Majumdar:1947eu,Papapetrou:1948jw}, which is also non-singular on and outside the (disconnected) event horizon~\cite{Hartle:1972ya} and unique~\cite{Chrusciel:1994qa,Heusler:1996ex}, but singular inside each of the black holes. Everywhere regular solitons, on the other hand, are again impossible, at least as \textit{static} solutions~\cite{Heusler:1996ft} or, more generically, \textit{strictly stationary} solutions~\cite{Shiromizu:2012hb}. Thus, at least as strictly stationary energy lumps, electro-vacuum does not realise particle-like solutions, that Wheeler dubbed \textit{geons} (gravitational electromagnetic entities)~\cite{Wheeler:1955zz}. 

The quest for self-gravitating solitons, thus, leads us beyond electro-vacuum. Historically, two developments may be highlighted. Firstly, Kaup~\cite{Kaup:1968zz} (see also~\cite{Ruffini:1969qy}) found a concrete realisation of ``geons" but in Einstein-(complex, massive)-Klein-Gordon theory. These solitons are now known as boson stars~\cite{Schunck:2003kk}. Secondly, Bartnik and McKinnon showed solitons also exist in Einstein-Yang-Mills theory~\cite{Bartnik:1988am} - see also the discussions in~\cite{Gibbons:1990um,Bizon:1994dh}. In both these cases, the existence of solitons is accompanied by the existence of ``hairy" black holes, $i.e.$ black holes that have macroscopic degrees of freedom not associated to a Gauss law - see~\cite{Herdeiro:2015waa,Volkov:2016ehx} for recent reviews. In the case of Einstein-Yang-Mills theory these are known as coloured black holes~\cite{Bizon:1990sr}; in the case of the Einstein-(complex, massive)-Klein-Gordon model these are called black holes with synchronised hair~\cite{Herdeiro:2014goa}. As a rule of thumb, one observes that in models in which both solitons and the standard Schwarzschild/Kerr black holes exist,  so does a (non-linear) bound state of both, which is a possible interpretation of the corresponding hairy black holes. But subtleties exist. For instance, in the Klein-Gordon case, the hairy black holes require rotation and do not exist in spherical symmetry~\cite{Pena:1997cy}. Turning around the rule of thumb, one may wonder if in a model where both ``bald" and hairy black holes exist, solitons should equally be found. In this paper, we shall illustrate this is not necessarily the case. 

In this work we shall consider an Einstein-Maxwell-scalar model, wherein the scalar field $\Phi$ is real and ungauged, has a canonical kinetic term and is non-minimally coupled to the Maxwell field via a coupling function $f(\Phi)$. The scalar field dynamics can thus be thought of as triggering a spacetime varying electric permittivity, controlled by $f(\Phi)$. The action is
\beq
\label{MODEL}
\mathcal{S}_{EMS} = \mathcal{S}_{EH} + \int d^4x\sqrt{-g}\bigg[- \frac{f(\Phi)}{4}F_{\mu\nu} F^{\mu\nu}-\frac{1}{2}\nabla^\mu\Phi\nabla_\mu\Phi-U(\Phi) \bigg]  \ .
\eeq
This model (without the potential energy $U(\Phi)$) was recently considered in~\cite{Herdeiro:2018wub} (see also the follow up works~\cite{Myung:2018vug,Boskovic:2018lkj,Myung:2018jvi,Fernandes:2019rez}) to address the spontaneous scalarisation of the Reissner-Nordstr\"om (RN) black holes.\footnote{See~\cite{Cardoso:2013opa,Cardoso:2013fwa} for spontaneous scalarisation of black holes triggered by matter and~\cite{Doneva:2017bvd,Silva:2017uqg,Antoniou:2017acq,Antoniou:2017hxj,Blazquez-Salcedo:2018jnn,Doneva:2018rou,Minamitsuji:2018xde,Silva:2018qhn,Brihaye:2019kvj} for scalarisation triggered by higher curvature couplings.} In~\cite{Herdeiro:2018wub} it was shown that under the condition $\frac{df}{d\Phi}(0)=0$, which may be interpreted as due to a $\mathbb{Z}_2$ symmetry, $\Phi\rightarrow -\Phi$, this model accommodates both the standard RN and new ``hairy" (or scalarised) black holes. Moreover, for sufficiently large charge to mass ratio, the RN black hole becomes unstable against scalarisation and fully non-linear numerical simulations~\cite{Herdeiro:2018wub} have shown the evolution of such instability saturates and forms a scalarised black hole, which is moreover perturbatively stable~\cite{Herdeiro:2018wub,Myung:2018jvi}, thus providing convincing evidence it is the end point of the instability.  Then, considering the foregoing discussion, it becomes interesting to ask whether, or under which conditions, this model may accommodate also self-gravitating solitons. In this paper we shall establish three no-go results for such particle-like solutions in particular cases of this model.

Firstly,  we set the Maxwell field to zero. This is a consistent truncation of the full model. In this case, using a scaling argument of the sort pioneered by Derrick to exclude stable solitons in a large class of (Minkowsi spacetime) field theories~\cite{Derrick:1964ww}, it is established that no stationary and axisymmetric self-gravitating scalar solitons exist, unless the scalar potential energy is somewhere negative in spacetime. This generalises previous results for the static~\cite{HeuslerBook,Heusler1995a} and strictly stationary cases~\cite{Shiromizu:2012hb} (the latter for vanishing potential, but allowing a negative cosmological constant). Thus, rotation alone cannot support self-gravitating scalar solitons in Einstein-(real-)scalar models.\footnote{One should note that there exist, however, quasi-stationary (indeed quasi-static) self-gravitating solitons in real scalar models with a mass term or more complicated positive potentials, named~\textit{oscillatons}~\cite{Seidel:1991zh}. Albeit, strictly speaking, non-static, these can be very long lived~\cite{Page:2003rd,Fodor:2009kg,Grandclement:2011wz}.}  Secondly, constant sign couplings are considered. Generalising a previous argument by Heusler for electro-vacuum~\cite{Heusler:1996ft}, it is established that no \textit{static} self-gravitating electromagnetic-scalar solitons exist, regardless of the spatial symmetries of the model. Thus, a varying (but constant sign) electric permittivity alone cannot support static Einstein-Maxwell solitons. The static case is already relevant for the questions raised before, as the scalarised black holes found in~\cite{Herdeiro:2018wub} are static and the coupling function used therein is sign invariant. Thus, the results herein, show they cannot be interpreted as a bound state of a RN black hole and a particle-like solution. Nonetheless, as our final result we are able to go beyond staticity and establish a similar absence of solitons for \emph{strictly stationary} spacetimes. This relies on the use of a Lichnerowicz-type argument, employed in~\cite{Shiromizu:2012hb}  for Einstein-Maxwell-scalar models, but wherein the scalar and Maxwell fields are not directly coupled. While a first version of the argument cannot accommodate a non-constant scalar potential, this is possible in a second version, influenced by~\cite{Heusler1995a, HeuslerBook}, and using the Komar mass integral. The scope of validity of each of these results points out possible paths to circumvent them, in order to obtain self-gravitating solitons in Einstein-Maxwell-scalar models.

This paper is organised as follows. In Section~\ref{section2} we establish the absence of stationary solitons for vanishing Maxwell field within   the model described by eq.~\eqref{MODEL}. Firstly, we show that the most generic line element is described by four metric functions. Then, we perform a scaling argument to show the scalar field must be trivial. A comparison with an argument coming from considering the scalar field equation but yielding a slightly different results is also made. In Section~\ref{section3} we establish the absence of static solitons in the model~\eqref{MODEL} without making any assumptions on spatial symmetries. In Section~\ref{section4} we generalise the previous argument for strictly stationary spacetimes. Finally, in Section~\ref{section5} conclusions and some final remarks are presented.

\section{Absence of stationary solitons for vanishing Maxwell field}
\label{section2}

The first result to be established concerns the absence of asymptotically flat, stationary and axisymmetric self-gravitating scalar solitons. Thus we consider $F_{\mu\nu}=0$. Observe this is a consistent truncation of the model~\eqref{MODEL}. That is, taking a vanishing Maxwell tensor in the action is equivalent to taking a vanishing Maxwell tensor in the field equations.  Thus, the model under consideration in this section is 
\beq\label{ActionPhi}
\mathcal{S} = \mathcal{S}_{EH} + \int d^4x \sqrt{-g}\bigg[-\frac{1}{2}\nabla^\mu\Phi\nabla_\mu\Phi-U(\Phi)\bigg] .
\eeq
Our first task is to show the most general metric form for the configurations we seek to rule out is
\beq
\label{gsm}
ds^2 = -\frac{\rho^2}{X(\rho,z)}dt^2 + X(\rho,z)\left[d\varphi - w(\rho,z) dt\right]^2 + A(\rho,z)\left[d\rho^2 + B(\rho,z)dz^2\right] \ ,
\eeq
which contains four unknown functions of  the ``cylindrical" coordinates $\rho$ and $z$. The second task is to rule such non-trivial solitonic solutions by applying a scaling argument.

\subsection{Most general line element}

\subsubsection{Isometries}

Firstly, axisymmetry and stationarity implies the existence of two Killing vector fields $m$ and $k$. Without loss of generality, these Killing vectors commute $[k, m]=0$~\cite{Carter:1970ea}, assuming the spacetime is asymptotically flat. Thus coordinates adapted simultaneously to \textit{both} these vectors fields can be chosen. As $k$ corresponds to the asymptotically timelike Killing vector field and $m$ to the spacelike one,  a temporal coordinate $t$ and an angular coordinate $\varphi$ are introduced along the orbits of the Killing vector fields as:
\beq 
k=\frac{\partial}{\partial t} \ , \;\;\;\;\; m = \frac{\partial}{\partial \varphi}\ .
\eeq
Consequently, in  coordinates $(t, \varphi, x, y)$, the general line element can be cast in the form:
\beq
ds^2 = g_{\mu\nu}(x,y)dx^\mu dx^\nu \ .
\eeq

\subsubsection{Circularity}

We now want to prove that our metric is \emph{circular}. That is, the surfaces orthogonal to the Killing fields $k$ and $m$ are integrable. By Frobenius' theorem (see~\cite{Wald:1984rg} App. B.3), the surfaces orthogonal to the Killing fields are integrable if the following conditions hold:
\beq
dk\wedge k\wedge m = 0 = dm\wedge m \wedge k  \ .
\eeq
Circularity means that the spacetime manifold $\mathcal{M}$ is locally a product of two 2-dimensional manifolds $\mathcal{M} = \mathcal{N}_1\times \mathcal{N}_2$ and can be cast in the following form:
\beq
ds^2 = g_{\mu\nu}(x,y)dx^\mu dx^\nu = \sigma_{ab}(x,y)dx^a dx^b + \gamma_{ij} (x,y) dx^idx^j \ ,
\eeq
where $\bm{\sigma}$ corresponds to the metric in the $(t,\varphi)$ manifold and $\bm{\gamma}$ corresponds to the metric in the $(x,y)$ manifold. Establishing circularity requires using the Einstein equations and hence depends on the energy-momentum of the spacetime. Circularity can actually be established by first establishing \textit{Ricci circularity} as we now discuss.

For the case under consideration, the energy-momentum tensor obtained from~\eqref{ActionPhi} is
\beq\label{TRS}
T_{\mu\nu} = \nabla_\mu\Phi\nabla_\nu\Phi -g_{\mu\nu}\bigg(\frac{1}{2}\nabla_\alpha\Phi\nabla^\alpha\Phi + U(\Phi)\bigg) \ .
\eeq
It follows that 
\beq
T(k)\wedge k \wedge m = 0 = T(m)\wedge m \wedge k \label{CircT} \ ,
\eeq
where the $T(k)$ and $T(m)$ 1-forms correspond to the contraction of the energy-momentum tensor with the Killing vectors. To establish this observe that, since the spacetime is stationary, the Einstein equations imply that $\pounds_k g_{\mu\nu} = 0 \Rightarrow \pounds_k T_{\mu\nu}=0$ and this in turn implies, due to $\Phi$ being real, that $\pounds_k \Phi= k^\mu\nabla_\mu\Phi=0$. Thus, 
\beq
T(k)_\mu = T_{\mu\nu}k^\nu = -k_\mu\bigg(\frac{1}{2}\nabla_\alpha\Phi\nabla^\alpha\Phi + U(\Phi)\bigg) \ ,
\eeq
meaning that $T(k)$ is proportional to $k$ and, as such, $T(k)\wedge k=0$. A similar argument shows that $T(m)\wedge m =0$, proving the equality \eqref{CircT}. Now, using Einstein's equations it follows that 
\beq
R(k)\wedge k \wedge m = 0 = R(m)\wedge m \wedge k \ ,
\label{Riccic}
\eeq
where the $R(k)$ and $R(m)$ 1-forms correspond to the contraction of the Ricci tensor with the Killing vectors. A spacetime which respects~\eqref{Riccic} is called \emph{Ricci circular}. Thus, we have shown that an asymptotically flat, axisymmetric and stationary spacetime sourced by a real scalar field (with an arbitrary potential) is Ricci circular. But Ricci circularity and circularity are equivalent for asymptotically flat, stationary and axisymmetric spacetimes~\cite{HeuslerBook,Carter:1969zz,Kundt:1966zz}, concluding the proof of circularity.\footnote{This proof can be extended for the full model \eqref{ActionPhi} including the electromagnetic field. Such extension, however, is not relevant for our discussion so we will omit the details regarding the electromagnetic field part of the action which can be found in \cite{HeuslerBook, Carter:1969zz}.} It follows we can then write the line element as
\beq\label{gW}
ds^2 = -Vdt^2 + 2Wdtd\varphi +  Xd\varphi^2 + \gamma_{ij}dx^idx^j \ ,
\eeq
where $V= -\langle k | k\rangle$, $X = \langle m | m\rangle$ and $W= \langle k | m\rangle$. We have now reduced the unknown metric functions from ten to six. 

Two remarks are in order. Firstly, observe that circularity is equivalent to assuming  the spacetime to be invariant under the simultaneous discrete symmetry transformations $(t,\varphi)\rightarrow (-t,-\varphi)$. However, the circularity argument shows that (for our matter content) this is not an assumption and no generality is lost. Secondly, for a \emph{complex} scalar field, circularity is actually lost. Indeed, we cannot guarantee the implication $\pounds_k T_{\mu\nu} = 0 \Rightarrow \pounds_k \Phi = 0$, as the field can have a harmonic dependence on both $t$ and $\phi$ in the form of a phase. In spherical coordinates the field would be ($n\in \mathbb{Z}$ and $w\in\mathbb{R}$ are constants)
\beq
\Phi(t, r, \theta, \varphi) = \phi(r,\theta)e^{i(n\varphi - \omega t)} \ ,
\eeq
and this dependence does not manifest itself in the energy-momentum tensor, which is still preserved independently by the two Killing vector fields.
Thus, the form~\eqref{gW} is no longer the most general metric form describing an asymptotically flat, axisymmetric, stationary spacetime sourced by a complex scalar field.\footnote{We thank E. Ay\'on-Beato for this observation.} Interestingly, this complex scalar field case allows circumventing no-scalar hair theorems for black holes, $e.g.$~Bekenstein's theorem \cite{Bekenstein:1972ny}, and yields black holes with scalar hair that are asymptotically flat, stationary and axisymmetric~\cite{Herdeiro:2014goa,Herdeiro:2015gia}. The known solutions have a geometry invariant under  $(t,\varphi)\rightarrow (-t,-\varphi)$; the absence of circularity opens up the possibility that more general hairy black holes may exist in the complex scalar field model.

\subsubsection{The orthogonal manifold}
The simplification of the orthogonal $(x,y)$ manifold, with metric $\gamma_{ij}$, can now be addressed. Due to the gauge freedom, $i.e.$  the ability to redefine the $(x,y)$ coordinates, one anticipates the possibility of reducing further the number of unknown functions from six to four. Indeed, one first introduces a scalar function  $\rho$,  defined as
\beq
\rho \equiv \sqrt{-\sigma} = \sqrt{VX+W^2} \ ,
\eeq
where $\sigma$ corresponds to the determinant of the $\sigma_{ab}$ metric. Assuming that both $\rho$ and $\nabla_\mu \rho$ do not vanish (to be further discussed below), we can choose $\rho$ as a coordinate on the orthogonal manifold. Introducing a second coordinate therein, $z$,  chosen such that $\nabla_\mu z\nabla^\mu\rho=0$ (by setting $z=$ constant along the integral curves of $\nabla^\mu\rho$), we can write the full metric in the form~\eqref{gsm}, where
\beq
w\equiv - \frac{W}{X} \ . 
\eeq
There are now four unknown metric components, $X,w, A$ and $B$ of two variables $\rho$ and $z$. Considering the way $\rho$ and $z$ were defined, this coordinate system is valid for $\rho\in]0,\infty[$ and $z\in]-\infty,\infty[$. We have now reached the maximal possible simplicity for the line element under our assumptions and for the matter content we wish to consider. Nonetheless, it is instructive to consider a further a simplification, which, however, is non-generic for our model. 

A further simplification can be made if $\rho$ is harmonic on the 2-dimensional orthogonal manifold, $\Box_{(\gamma)}^2\rho=0$. In this case, $z$ will be the harmonic conjugate function of $\rho$ and they will obey the Cauchy-Riemann equations, which give us the following expressions:
\begin{align}
\gamma_{\rho z} = \langle \nabla_{(\gamma)} \rho &| \nabla_{(\gamma)} z \rangle = 0 \ , \\
\gamma_{\rho\rho} = \langle \nabla_{(\gamma)} \rho | \nabla_{(\gamma)} \rho \rangle &= \langle \nabla_{(\gamma)} z | \nabla_{(\gamma)} z \rangle = \gamma_{zz} \ .
\end{align}
We can then set $B=1$ in \eqref{gsm} and, after redefining $A\equiv e^{2h}/X$ (for consistency with the literature), we obtain the Weyl-Lewis-Papapetrou (WLP) metric:
\beq\label{WLP}
ds^2 = -\frac{\rho^2}{X(\rho,z)}dt^2 + X(\rho,z)\left[d\varphi - w(\rho,z) dt\right]^2 + \frac{e^{2h(\rho,z)}}{X(\rho,z)}\left[d\rho^2 + dz^2\right] \ ,
\eeq
which has only three unknown functions.

Another important consequence of $\rho$ being harmonic is that, if $\rho$ is not a constant, it can be shown that it has \emph{no critical points} in the orthogonal manifold (no points where the gradient vanishes)~\cite{CarterBHE}. As this choice of coordinates is well behaved except for when $d\rho=0$, it means the coordinate system is globally well behaved in the whole manifold, as long as there are no event horizons. So the metric \eqref{WLP} can be used to describe the whole stationary and axisymmetric spacetime.

The key question for this further simplification is: when can it be guaranteed that $\rho$ is harmonic? It has been proven by Papapetrou and others~\cite{Papapetrou:1966zz,HeuslerBook} that $\rho$ is harmonic as long as the projection of the Ricci tensor along the $(t,\varphi)$ surfaces is trace free through the following equation \cite{HeuslerBook}:
\beq
\frac{1}{\rho}\Box^2_{(\gamma)}\rho = -\frac{1}{X}tr_\sigma\bm{R} \ .
\eeq
 Using the metric \eqref{gW} for simplicity, this translates into the following condition for $\rho$ to be harmonic:
\beq
tr_{\sigma}\bm{R} = \sigma^{ab}R_{ab} = \frac{1}{\rho^2}\big[-XR(k,k) + 2WR(k,m) + VR(m,m)\big]=0 \ .
\label{harmonicc}
\eeq
So as long as~\eqref{harmonicc} is respected, $\rho$ is harmonic and the coordinates used in the metric \eqref{WLP} are globally well defined: the $\gamma_{ab}$ metric is \emph{globally conformally flat}.\footnote{Every two-dimensional metric is \emph{locally conformally flat} due to the existence of isothermal coordinates (the uniformization theorem \cite{UniTh1,UniTh2}). Such choice of coordinates is, however, only \emph{locally} and not \emph{globally} conformally flat. Thus one cannot guarantee their validity throughout the whole orthogonal manifold. Invoking such conformal flatness would reduce the unknown metric functions from six to four, instead of the three obtained in the WLP metric, since $\rho$ would not be used as a coordinate any longer.} Now all that is left is to check is if our model respects condition~\eqref{harmonicc}. We can write the Ricci tensor in terms of the energy momentum tensor as follows $8\pi G=c=1$):
\beq
R_{\mu\nu} = \bigg(T_{\mu\nu} - \frac{1}{2}g_{\mu\nu}T \bigg) \ .
\eeq
The Ricci tensor components $R(k,k)$, $R(k,m)$ and $R(m,m)$ for the real scalar field energy momentum tensor \eqref{TRS} read
\begin{align}
&R(k,k) = R_{\mu\nu}k^\mu k^\nu = - U(\Phi) V \ , \\
&R(m,m) = R_{\mu\nu}m^\mu m^\nu =  U(\Phi) X \  ,\\
&R(k,m) = R_{\mu\nu}k^\mu m^\nu =  U(\Phi) W \ ,
\end{align}
yielding
\beq
tr_{\sigma}\bm{R} =  2 U(\Phi)\ .
\eeq
This can only be zero in the whole manifold if the potential $U(\phi)$ vanishes. So the metric \eqref{WLP} is the most general metric for a free real scalar field only; for a non-zero potential we must resort to the form~\eqref{gsm}, as we will do in the next subsection.

\subsection{Scaling argument}
We have established that the most general line element that can describe the solitonic solutions we seek is given by eq. \eqref{gsm}, 
where $(t,\rho,\varphi,z)$ are ``cylindrical" coordinates and $A, B, w$ and $X$ are unknown functions of the non-Killing coordinates ($\rho,z$) of which $A$, $B$ and $X$ are all positive. We can now proceed to show that there are no non-trivial solutions of this form for the model~\eqref{ActionPhi} by a scaling argument. It is useful to observe that the square root of minus the metric determinant is
\beq
\sqrt{-g} = \rho A\sqrt{B} \label{detg} \ .
\eeq

We start with the Einstein-Klein-Gordon action for the scalar field \eqref{ActionPhi}. The proof follows by contradiction. We assume that such a stationary, axisymmetric, asymptotically flat self-gravitating scalar soliton exists. Because the scalar field is real, it respects, for this hypothesised solution the same symmetries as the metric possesses, so it does not change under the action of the stationary and axisymmetric Killing vectors. The next step is to consider a scaling of the hypothesised solution by a scale factor $\lambda$. This rescales the coordinates ($\rho$, $z$) and defines a one-parameter family of configurations (not necessarily solutions) of the coupled geometry-scalar system:
\begin{align}
A_\lambda(z,\rho) = A(\lambda z, \lambda\rho)\ , \;\;\;&\;\; B_\lambda(z,\rho) =B(\lambda z,  \lambda\rho)\ ,\;\;\;\;\;w_\lambda(z,\rho) = w(\lambda z, \lambda\rho)\ ,\\
X_\lambda(z,\rho) = &X(\lambda z, \lambda\rho)\ , \;\;\;\;\;\;\Phi_\lambda(z,\rho) = \Phi(\lambda z, \lambda\rho)\ .
\end{align}
Under this scaling transformation, the metric determinant transforms as, using \eqref{detg},
\beq\label{trf1}
\int d^4x \sqrt{-g}  \rightarrow  \int d^4x_\lambda \sqrt{-g_\lambda} = \lambda^3 \int d^4x\, \rho A_\lambda\sqrt{B_\lambda} \ ,
\eeq
and the kinetic scalar field term transforms as
\beq\label{trf2}
-\nabla^\mu\Phi\nabla_\mu\Phi = -\frac{1}{A}\bigg[ (\partial_\rho\Phi)^2 + \frac{1}{B}(\partial_z\Phi)^2\bigg] \rightarrow -\frac{1}{\lambda^2 A_\lambda}\bigg[ (\partial_\rho\Phi_\lambda)^2 + \frac{1}{B_\lambda}(\partial_z\Phi_\lambda)^2\bigg] \ .
\eeq
The action of the scaled solutions is $\mathcal{S}^\lambda = \mathcal{S}[\Phi_\lambda, B_\lambda, A_\lambda, w_\lambda, X_\lambda]$. Since for $\lambda=1$ we have the hypothesised solution, the variation of $\mathcal{S}^\lambda$  with respect to $\lambda$ must have a stationary point at $\lambda=1$. This condition yields the virial relation:\footnote{Only the matter field action enters the argument because the Einstein-Hilbert part of the action is invariant under a rescaling transformation as it corresponds to a \emph{diffeomorphism}. In other words, we have that $\delta \mathcal{S}^\lambda_{EH}/\delta\lambda = 0$.}
\beq
\frac{1}{2}\int_0^{+\infty} \!\!\!\!\!d\rho\int_{-\infty}^{+\infty}\!\!\!\!\! dz\, \rho \bigg[\sqrt{B} (\partial_\rho\Phi)^2 + \frac{1}{\sqrt{B}}(\partial_z\Phi)^2\bigg] = -3\int_0^{+\infty}\!\!\!\!\! d\rho\int_{-\infty}^{+\infty}\!\!\!\!\! dz \,\rho A\sqrt{B}\,\, U(\Phi)\ .
\label{VirialS} 
\eeq
As the left side is always non-negative, and the right hand side is always non-positive, for positive $U(\Phi)$ we get to a contradiction, which can only be settled if the hypothesised solution is trivial. Alternatively, a negative potential is mandatory, to have a non-trivial solitonic solution. Thus, the only asymptotically flat, everywhere regular, stationary and axisymmetric localized solution for the model \eqref{ActionPhi} with $U(\Phi)\geqslant 0$ is Minkowski spacetime. No self-gravitating scalar solitons exist in this model and under these assumptions.

The virial identity~\eqref{VirialS} can be written in a more compact form as 
\beq\label{VirialS2}
\int d^4x\sqrt{-g}\bigg[\nabla^\mu\Phi\nabla_\mu\Phi + 6 U(\Phi)\bigg] =  0 \ .
\eeq
One observes the resemblance with the identity that Bekenstein deduced when attempting to rule out a black hole spacetime with scalar hair in the same model~\cite{Bekenstein:1972ny}:
\beq\label{BekEq}
\int d^4x\sqrt{-g}\bigg[\nabla^\mu\Phi\nabla_\mu\Phi + \Phi \frac{d U(\Phi)}{d\Phi}\bigg]=0 \ .
\eeq
To obtain the latter identity for the model~\eqref{ActionPhi} one integrates the Klein-Gordon equation $\nabla^\mu\nabla_\mu\Phi - U'(\Phi)= 0$ multiplied by $\Phi$ over the whole spacetime and then, upon integrating by parts the first term, a surface term at infinity emerges, which vanishes since $\Phi\nabla_\mu\Phi\rightarrow 0$ at infinity for an asymptotically flat spacetime. This procedure yields~\eqref{BekEq}.\footnote{Another slight variation consists on multiplying the Klein-Gordon equation by  $dU/d\Phi$ instead of $\Phi$~\cite{Sotiriou:2011dz}. In this case one gets an obstruction under the condition of the convexity of the potential, $d^2U/d\Phi^2>0$ - see also~\cite{Herdeiro:2015waa}.}

A key difference between the Bekenstein identity~\eqref{BekEq} and the virial identity \eqref{VirialS2} is that the latter is independent from the equations of motion, while the former is a consequence of the scalar equation of motion. Moreover, they yield different (but complementary) conclusions. In particular, the Bekenstein identity with a positive potential is not enough to obtain the no go theorem we have just described. Rather it would rule out gravitating solitons under the assumption that $\Phi U'(\Phi)>0$ everywhere (except for some discrete points where it can vanish), rather than the positivity of the potential. 

Another remark concerns the case of a constant, but non-zero, potential $U(\Phi)=\Lambda$, that can be interpreted as a cosmological constant. Does  the virial identity \eqref{VirialS} encode a no-go theorem for free spinning solitons in de Sitter spacetime ($\Lambda>0$)? Actually no, since for de Sitter spacetime the metric \eqref{gsm} is not necessarily the most general metric. This is because the de Sitter spacetime is not asymptotically flat, a requirement to guarantee both that the commutativity of the Killing vectors \cite{Carter:1970ea} and that the circularity theorem~\cite{HeuslerBook,Carter:1969zz,Kundt:1966zz} holds. Moreover, of course, de Sitter is not stationary, and the very definition of an equilibrium self-gravitating soliton has to be reconsidered. In the case of Anti-de-Sitter, no conclusions can be inferred either, but we remark that a no-go theorem for self-gravitating, purely gravitational solitons in Anti-de-Sitter was presented in~\cite{Boucher:1983cv}.

\section{Absence of static scalar-electromagnetic solitons}
\label{section3}
We now turn to the full model~\eqref{MODEL} to rule out \textit{static}, asymptotically flat scalar-electromagnetic solitons. In this case no spatial symmetry assumption is made. The argument generalises an electro-vacuum argument by Heusler~\cite{Heusler:1996ft}.

\subsection{Heusler's argument for static spacetime}
\label{section31}
Our focus in this subsection is a static spacetime. We consider an asymptotically flat, everywhere regular and static spacetime with a strictly stationary Killing field $k$, obeying $V\equiv -k^\mu k_\mu$>0, and shall prove that there are no solitons in the full model theory. The Einstein-Hilbert action $\mathcal{S}_{EH}$ will not play a role in this argument and neither will the metric as only the electromagnetic field equations of motion are used.

Define the electric and magnetic fields as:
\begin{align}
E_\mu &= -F_{\mu\nu}k^\nu \label{efield}\ , \\
B_\mu &= -\frac{1}{2}\varepsilon_{\mu\alpha\beta\nu} F^{\alpha\beta} k^\nu \label{bfield} \ ,
\end{align}
where $\varepsilon_{\mu\alpha\beta\nu}$ is the Levi-Civita tensor. The Maxwell equations for the full model are
\begin{align}
&\nabla_{[\mu}E_{\nu]}= 0\ , \\
&\nabla_{[\mu}\big(fB_{\nu]}\big) = 0 \ , \\
&\nabla_\mu\left(f\frac{E^\mu}{V} \right) = 0 \ , \label{Max3}\\
&\nabla_\mu\left(\frac{B^\mu}{V} \right) = 0 \ . \label{Max4}
\end{align}
Due to the absence of currents, the first two Maxwell equations imply that the electric $\varphi$ and magnetic-like $\psi$ \textit{scalar} potentials can be introduced, as $E_\mu=\partial_\mu\varphi$ and $fB_\mu=\partial_\mu\psi$.
A general mathematical identity states that for an arbitrary vector $\alpha$ that respects $\pounds_k\alpha=[k,\alpha]=0$:
\beq
\int_{\partial\Sigma}\alpha^{\mu}k^{\nu}dS_{\mu\nu} = \frac{1}{2}\int_{\Sigma} \nabla_\mu\alpha^\mu k^\nu d\Sigma_\nu\label{Stokes} \ ,
\eeq
where $\Sigma$ is an hypersurface, with volume element $d\Sigma_\nu$, while $\partial\Sigma$ corresponds to its boundary, with antisymmetric area element $dS_{\mu\nu}$. This is a version of Stokes' theorem in the presence of a Killing field. Applying this identity with $\alpha^\mu \rightarrow fE^\mu/V$ we obtain, using Maxwell's equations, 
\beq
\int_{\partial\Sigma}f\frac{E^{\mu}k^{\nu}}{V}dS_{\mu\nu} = 0\ .
\eeq
If $\Sigma$ is any Cauchy surface, we can take $\partial\Sigma$ to be the surface at infinity, in which case:\footnote{Here we are assuming there is no boundary term at the origin. This would not be the case if the coupling diverges at the origin, with an appropriate power of $1/r$. Then, there could be a boundary term at the origin. Thus, we require the coupling to be everywhere finite, $0<f(\Phi)<\infty$.}
\beq
\int_{\partial\Sigma}f\frac{E^{\mu}k^{\nu}}{V}dS_{\mu\nu} = -4\pi Q_e f_\infty = -4\pi Q_e = 0 \ ,
\eeq
where $Q_e$ corresponds to the electric charge. If we now replace $\alpha^\mu$ by $\varphi fE^\mu/V$ and once again use the Maxwell equations, we obtain
\beq
\frac{1}{2}\int_{\Sigma}f\frac{E^\mu E_\mu}{V}k^\nu d\Sigma_\nu = \int_{\partial\Sigma}\varphi f\frac{E^{\mu}k^{\nu}}{V} dS_{\mu\nu} = -4\pi Q_e\varphi_\infty = 0 \ .
\eeq
Since $k^\mu E_\mu = 0$ then $E$ is never timelike and, assuming that the coupling $f$ does not change sign, it follows that this expression only holds if the electric field vanishes. The same argument can be used for $B$ by replacing $\varphi$ by $\psi$, obtaining
\beq
\int_{\Sigma}f\frac{B^\mu B_\mu}{V}k^\nu d\Sigma_\nu = 0 \ ,
\eeq
from which we conclude that $B$ must also vanish if $f$ does not change sign. Thus, for a constant sign coupling function, solitons with a non-trivial electromagnetic field are ruled out, regardless of the potential $U(\Phi)$.

It remains the possibility that there could be self-gravitating solitons with a non-trivial scalar field. However, the scalar field must also vanish, as long as it obeys the dominant energy condition \textit{and} violates the strong energy condition~ \cite{HeuslerBook,Heusler1995a}. The rationale is the following. If the scalar field violates the strong energy condition this implies its Komar mass $M$ is negative; but if it respects the dominant energy condition, the positive mass theorem is applicable and its ADM (or Komar) mass is non-negative. This leads us to a contradiction. So it remains to see what these conditions mean for the model. Consider the energy-momentum tensor of the full action~\eqref{MODEL}:
\beq\label{EMTensor}
T_{\mu\nu} = \nabla_\mu\Phi\nabla_\nu\Phi -g_{\mu\nu}\bigg(\frac{1}{2}\nabla_\alpha\Phi\nabla^\alpha\Phi + U(\Phi)\bigg) + f(\Phi)\bigg(F_{\mu\alpha}F_\nu^\alpha-\frac{1}{4}g_{\mu\nu}F_{\alpha\beta}F^{\alpha\beta}\bigg) \ .
\eeq
The strong energy condition requires $R_{\mu\nu}\tilde{k}^\mu \tilde{k}^\nu\geqslant 0$ for any timelike vector field $\tilde{k}^\mu$ (for instance one that obeys $\tilde{k}^\mu=k^\mu/\sqrt{V}$). For $F=0$, and a static spacetime with a purely spatial scalar field distribution, this yields, 
\beq
U\leqslant 0 \ .
\eeq
Thus, a scalar field with a non-negative potential only obeys the strong energy condition if the potential is trivial. If, moreover, it obeys the dominant energy condition, then the aforementioned contradiction applies, except if $U=0$. In that case, the Komar mass is zero, and by the positive energy theorem, the resulting solution is Minkowski spacetime.

Since the result in this section relies on the constancy of the sign of $f(\Phi)$, one may ask what is the physical meaning of a change in the sign of the coupling function $f(\Phi)$. To assess this, observe that from  the energy-momentum tensor of the full action~\eqref{EMTensor}, the energy density is given as
\beq
\rho = T_{\mu\nu}k^\mu k^\nu = V\bigg(\frac{1}{2}\nabla_\alpha\Phi\nabla^\alpha\Phi + U(\Phi)\bigg) + f(\Phi)\bigg(E_\alpha E^\alpha+\frac{V}{4} F_{\alpha\beta}F^{\alpha\beta}\bigg) \ ,
\eeq
where $V=-k^\mu k_\mu$ is always positive as the spacetime is strictly stationary. We see that the energy density contribution of the electromagnetic field will, generically,  change sign along with $f(\Phi)$. Another perspective is that the electric permittivity would also change sign. Both these observations make such sign change physically questionable, as it would make the electromagnetic contribution a sort of exotic matter. It is worth noting, however, that the weak energy condition $\rho>0$ needs not be violated even if such sign change in the coupling occurs, as the scalar field contribution could compensate for the opposite sign contribution of the electromagnetic field. 

To close this section, let us remark on the dominant energy condition. In the full model, if the coupling $f(\Phi)$ changes sign to negative, then the full model may not respect the dominant energy condition even if, separately, the electromagnetic and scalar parts (excluding the coupling) abide it. This can be seen as follows: the dominant energy condition states that
\beq
T_{\mu\nu}X^\mu Y^\nu \geqslant 0 \ ,
\eeq
for any two co-oriented causal vectors $X^\mu$ and $Y^\mu$. Then, even assuming the scalar and electromagnetic EM tensors obey it
\begin{align}
T^{S}_{\mu\nu}X^\mu Y^\nu \geqslant 0 \ , && T^{E}_{\mu\nu}X^\mu Y^\nu \geqslant 0 \ ,
\end{align}
the full model EM tensor is $T_{\mu\nu} = T^{S}_{\mu\nu} + f(\Phi) T^{E}_{\mu\nu}$, which needs not respect the dominant energy condition due to the sign of $f(\Phi)$. While a negative $f(\Phi)$ does not directly imply that the model does not respect the dominant energy condition, this possibility ties with the fact that a different sign for $f(\Phi)$ implies that the electromagnetic field behaves like exotic matter. It seems, thus, that in the most reasonable physical scenarios, $f(\Phi)$ should not change sign and the dominant energy condition will hold for the full model as long as it obeys, separately, for the electromagnetic and scalar sectors.

\section{Absence of strictly stationary scalar-electromagnetic solitons}
\label{section4}

A further step beyond the last result  in the direction of generality, is to rule out \textit{strictly stationary}, but not necessarily static, asymptotically flat scalar-electromagnetic solitons in our full model~\eqref{MODEL}. With this goal in mind, we consider a Lichnerowicz-type argument adapting the one presented in \cite{Shiromizu:2012hb} where it was applied to Einstein-Maxwell-scalar models, but where the (possibly complex) scalar field has no direct coupling to the electromagnetic field. This argument consists in finding a divergence identity from which we may restrict the ADM mass of the system to vanish. Thus, as long as the dominant energy condition holds, one can conclude, from the positive mass theorem, that the spacetime is Minkowski. 

This argument generalises the one presented in last section in the sense it does not require staticity. Moreover, with respect to the argument in Section~\ref{section2}, it assumes an everywhere timelike Killing vector field (and hence an absence of ergo-regions) which is not a requirement in Section~\ref{section2}; in the latter, on the other hand, axial symmetry is assumed, unlike the argument here which has no spatial symmetry requirements.

\subsection{Lichnerowicz argument for strictly stationary spacetimes}

We consider an asymptotically flat, everywhere regular and strictly stationary spacetime with Killing field $k$. We can write the Einstein equations for our full model \eqref{MODEL} as
\beq\label{EinsteinEq}
R_{\mu\nu} = f(\Phi)\left(F_\mu^{\,\,\alpha}F_{\nu\alpha} - \frac{1}{4}g_{\mu\nu}F^2\right) + \partial_\mu\Phi\partial_\nu\Phi + g_{\mu\nu}U(\Phi) \ .
\eeq
We define the \textit{twist vector} $\omega^\mu$ using the timelike Killing vector
\beq
\omega^\mu = \frac{1}{2}\varepsilon^{\mu\nu\alpha\beta}k_\nu\nabla_\alpha k_\beta \ ,
\label{twistvector}
\eeq
which respects the identity
\beq\label{TwistId}
\nabla_\mu\bigg(\frac{\omega^\mu}{V^2}\bigg)=0 \ ,
\eeq
where, as in Section~\ref{section2}, $V\equiv -k^\mu k_\mu$, and here it is assumed to be always positive, corresponding to a strictly stationary spacetime.

The electric and magnetic fields are defined as in \eqref{efield} and \eqref{bfield}. The Maxwell equations take the following form in a strictly stationary spacetime:
\begin{align}
&\nabla_{[\mu}E_{\nu]}= 0\ , \label{Max21} \\
&\nabla_{[\mu}\big(fB_{\nu]}\big) = 0 \ , \\
&\nabla_\mu\left(f\frac{E^\mu}{V} \right) = \frac{2}{V^2}f\omega_\mu B^\mu \ ,\\
&\nabla_\mu\left(\frac{B^\mu}{V} \right) = -\frac{2}{V^2}\omega_\mu E^\mu \ . \label{Max24}
\end{align}
Observe how dropping the staticity assumption generalises the last two equations, as compared to their counterparts in the static case~\eqref{Max3}-\eqref{Max4}. On the other hand, since the two first equations remain the same, we can, as before,  write the fields in terms of two potentials $E_\mu=\partial_\mu \phi$ and $fB_\mu = \partial_\mu \psi$.

Using the relation
\beq
\nabla_{[\mu}\omega_{\nu]} = \frac{1}{2}\varepsilon_{\mu\nu}^{\;\;\;\;\alpha\beta}k_{[\alpha}R_{\beta]\gamma}k^\gamma \ ,
\eeq
we can obtain through the Einstein equations \eqref{EinsteinEq} the following expression
\beq
\nabla_{[\mu}\omega_{\nu]} =  f B_{[\mu}E_{\nu]}  \ .
\eeq
Roughly, the curl of $\omega$ is the Poynting vector. Then, from the existence of the potentials $\phi$ and $\psi$, we can obtain two more equations
\begin{align}
\nabla_{[\mu}\left(\omega_{\nu]}-\psi E_{\nu]}\right) = 0 \ ,\\
\nabla_{[\mu}\left(\omega_{\nu]}+ f\phi B_{\nu]}\right) = 0 \ ,
\end{align}
which, in turn, imply the existence of two further scalar potentials
\begin{align}
\nabla_\mu U_E &= \omega_\mu-\psi E_\mu \ ,\\
\nabla_\mu U_B &= \omega_\mu+ f\phi B_\mu \ .
\end{align}
Using these equations, the Maxwell equations~\eqref{Max21}-\eqref{Max24} and equation \eqref{TwistId}, we can obtain the following two identities
\begin{align}
\nabla_{\mu}\left(U_E\frac{\omega^\mu}{V^2}-\frac{\psi}{2V}B^\mu\right)=\frac{\omega_\mu\omega^\mu}{V^2} -  f\frac{B_\mu B^\mu}{2V} \ ,\label{EqU1}\\
\nabla_{\mu}\left(U_B\frac{\omega^\mu}{V^2}-\frac{\phi}{2V}fE^\mu\right)=\frac{\omega_\mu\omega^\mu}{V^2} -  f\frac{E_\mu E^\mu}{2V}  \ .\label{EqU2}
\end{align}
Another useful identity is the contraction of the Ricci tensor \eqref{EinsteinEq} with the stationary Killing field twice
\beq\label{RiccKilling}
R_{\mu\nu}k^\mu k^\nu = \frac{f}{2}\left(E_\mu E^\mu + B_\mu B^\mu\right) - VU(\Phi) \ .
\eeq

We now consider two different approaches to further the argument.

\subsubsection{First approach}
This approach follows the one in~\cite{Shiromizu:2012hb} but including now the non-minimal coupling function $f(\Phi)$. 

Using the following relation for the twist vector (see for example~\cite{HeuslerBook})
\beq
\frac{2}{V}R_{\mu\nu}k^\mu k^\nu =\nabla_\mu\left(\frac{\nabla^\mu V}{V}\right) + 4\frac{\omega_\mu\omega^\mu}{V^2} \ ,
\eeq
together with equation \eqref{RiccKilling}, we get
\beq
\nabla_\mu\left(\frac{\nabla^\mu V}{V}\right) + 4\frac{\omega_\mu\omega^\mu}{V^2}  = \frac{f}{V}\left(E_\mu E^\mu + B_\mu B^\mu\right) - 2U(\Phi) \ .
\eeq
This relation together with \eqref{EqU1} and \eqref{EqU2} finally gives the divergence identity:
\beq\label{DivId}
\nabla_{\mu}\left[ \frac{\nabla^\mu V}{V} + 2(U_E + U_B)\frac{\omega^\mu}{V^2} - \frac{\psi B^\mu + f\phi E^\mu}{V}\right] = -2U(\Phi) \ .
\eeq

Let us now analyse the consequences of this divergence identity, starting with the particular case of a free, massless, scalar field, so that $U(\Phi)=0$. The left hand side of~\eqref{DivId} is the divergence of a vector $v^\mu$ which respects 
\beq
k^\mu v_\mu=0 \ ,
\label{kv}
\eeq
which follows from the definitions~\eqref{efield}, \eqref{bfield}, \eqref{twistvector} and $V\equiv -k^\mu k_\mu$. Now we integrate $\nabla_\mu v^\mu$ on a spacetime volume bounded by two (neighbouring) Cauchy hypersurfaces, $\Sigma_1$ and $\Sigma_2$, with exterior normals $k^\mu$ and $-k^\mu$, respectively, and  a timelike hypersurface at spatial infinity, $\mathcal{T}$, whose spatial sections are round 2-spheres, and hence the normal is a unit radial vector $n^\mu$. Then, applying the covariant divergence theorem:
\beq
0=\int_{V^4}d^4 x \sqrt{-g} \nabla_\mu v^\mu=\int_{\Sigma_1}d^3x\sqrt{g_\Sigma}k_\mu v^\mu-\int_{\Sigma_2}d^3x\sqrt{g_\Sigma}k_\mu v^\mu+\int_{\mathcal{T}}d^3x\sqrt{-g_\mathcal{T}}n_\mu v^\mu \ .
\eeq
The first two integrals in the right hand side vanish by~\eqref{kv}. To simplify the remaining term we note that, asymptotically, 
 the leading behavior of the asymptotically flat metric is
\beq
ds^2 = - Vdt^2 + \frac{dr^2}{V} + r^2(d\theta^2+\sin^2\theta d{\varphi}^2) + ...
\eeq
where
\beq
V = 1-\frac{2M}{r}+\mathcal{O}\left(\frac{1}{r^2}\right) \ .
\eeq
Then\footnote{The electromagnetic terms in the divergence \eqref{DivId} disappear at infinity because they all decay asymptotically faster than $r^{-2}$. Consequently, only the first term inside the divergence contributes to this integral.},
\beq
0=\int_{\mathcal{T}}d^3x\sqrt{-g_\mathcal{T}}n_\mu v^\mu= \lim_{r\rightarrow \infty}\int_{t_2}^{t_1} dt \int_0^{2\pi} d{\varphi}\int_0^{\pi} d\theta r^2\sin\theta \sqrt{V} n_r \frac{\partial_rV}{V}=8\pi  M \Delta t \ .
\eeq
where $\Delta t=t_1-t_2$ and $t_1,t_2$ are the (arbitrary) time coordinates associated to the two Cauchy surfaces. This informs that the ADM mass $M$ must vanish. Then, by the positive mass theorem, assuming the  dominant energy condition holds for this model, the spacetime is Minkowski.

Concerning the scalar potential, unless  $U(\Phi)$ is written as a divergence, as to be included in the left side of the equation \eqref{DivId}, the reasoning does not apply. This can be done for a constant negative potential.\footnote{The process of integration in this case is exactly the same as in \cite{Shiromizu:2012hb}.} But for the general case we shall follow a different approach.

\subsubsection{Second approach}

In order to accommodate a non-trivial potential in the no-go theorem, we now take advantage of an argument in~\cite{Heusler1995a}, already used in Section~\ref{section31}, stating that a strictly stationary, asymptotically flat spacetime coupled to a matter model satisfying the dominant energy condition will always be flat spacetime as long as it violates the strong energy condition for the Killing field $k$ at every point, $R_{\mu\nu}k^\mu k^\nu\leqslant 0$. 

We start from the following result in~\cite{Heusler1995a} for the Komar mass:
\begin{align}\label{KomarM}
M&= -2\int_\Sigma\left(\frac{R_{\mu\nu}k^\mu k^\nu}{V} - \frac{2\omega^\mu\omega_\mu}{V^2}\right)k^\alpha d\Sigma_\alpha \nonumber \\
&= 2\int_\Sigma\left(R_{\mu\nu}k^\mu k^\nu - \frac{2\omega^\mu\omega_\mu}{V}\right) d\Sigma \ ,
\end{align}
where $\Sigma$ is a spacelike Cauchy surface.\footnote{This means that $k^\alpha d\Sigma_\alpha = k^\alpha k_\alpha d\Sigma = - V d\Sigma$.} Since $\omega^\mu$ is nowhere timelike, $\omega^\mu\omega_\mu>0$ and, if $R_{\mu\nu}k^\mu k^\nu\leqslant 0$, both contributions to the integral will be negative and
\beq
M \leqslant 0 \ .
\eeq
Assuming the dominant energy condition, on the other hand,  the positive mass theorem $M\geqslant 0$ holds. Thus, $M=0$ and the spacetime is flat. 

For  our full model, we cannot simply state that the strong energy condition is violated, due to the presence of the electromagnetic field terms in equation \eqref{RiccKilling}. Nonetheless, we can still show the mass is non-positive, for a non-negative potential. We then proceed as follows. Using equations \eqref{EqU1}, \eqref{EqU2} and \eqref{RiccKilling} we obtain
\beq\label{RiccKilling2}
\int_{\Sigma}R_{\mu\nu}k^\mu k^\nu d\Sigma = \int_{\Sigma}\bigg(\frac{2\omega^\mu\omega_\mu}{V} - V\,U(\Phi) -\frac{V}{2}\nabla_\mu W^\mu\bigg) d\Sigma \ ,
\eeq
where we defined the vector $W^\mu$ as
\beq
W^\mu \equiv 2(U_E + U_B)\frac{\omega^\mu}{V^2} - \frac{\psi B^\mu + f\phi E^\mu}{V} \ .
\eeq
Using the full Komar expression \eqref{KomarM}, equation \eqref{RiccKilling2} becomes
\beq\label{KomarM2}
M = -2\int_\Sigma V\, U(\Phi) d\Sigma - \int_\Sigma V\,\nabla_\mu W^\mu d\Sigma \ .
\eeq
For a positive potential, the first term will be clearly negative so we only have to deal with the second term which corresponds to the electromagnetic field contribution to the Komar mass. First, note that $\pounds_k W = 0$ so we can use the Stokes theorem identity \eqref{Stokes} for this vector
\beq\label{SInt}
\int_\Sigma V\,\nabla_\mu W^\mu d\Sigma = -\int_\Sigma \nabla_\mu W^\mu k^\nu d\Sigma_\nu = -2\int_{\partial\Sigma}W^\mu k^\nu dS_{\mu\nu} \ .
\eeq
The surface $\partial\Sigma$ is the 2-surface at infinity and all the terms in $W^\mu$ decay, asymptotically, faster than $r^{-2}$, so the integral vanishes. This means that the electromagnetic contribution to the Komar mass \eqref{KomarM2} given by the vector $W^\mu$ is zero and only the negative potential term is left, giving, again, $M\leqslant 0$. Thus, again,  the positive mass theorem establishes $M=0$. 

To conclude, in the full model \eqref{MODEL}, in an asymptotically flat and strictly stationary spacetime with positive scalar potential $U(\Phi)$, there are no non-trivial solitonic solutions, even in the presence of any positive potential $U(\Phi)$.\footnote{We emphasise that in Section~\ref{section4} we have assumed, in the applications of Stokes' theorem, the absence of a boundary term at the origin, which, again, is only justified if the coupling is required to be finite therein. Singular solutions with divergent coupling have been reported, for instance, in~\cite{Monni:1995vu,Cadoni:2009xm}.} Observe that this generalises the argument of~\cite{Shiromizu:2012hb} even for the case of minimal coupling $f(\Phi)=1$, due to the inclusion of the potential.

\section{Conclusions and final remarks}
\label{section5}

In flat spacetime, solitonic solutions occur in classical field theories that are, just like general relativity, nonlinear. Examples date back as far as the Korteweg-de-Vries equation \cite{KdV}. Coupling field theories to gravity opens up new possibilities: (i) solitonic solutions could arise even for linear field theories, with the required non-linearities being generated by the gravitational interaction; (ii) there are now also black hole solutions, which in vacuum are ``bald". When both black hole solutions and solitonic solutions are possible, often so are a ``hairy" black hole solutions, which can be faced as a sort of non-linear bound state between both of these building blocks. 

	\begin{figure}[h]
		 \centering
	 	 \includegraphics[scale=0.4]{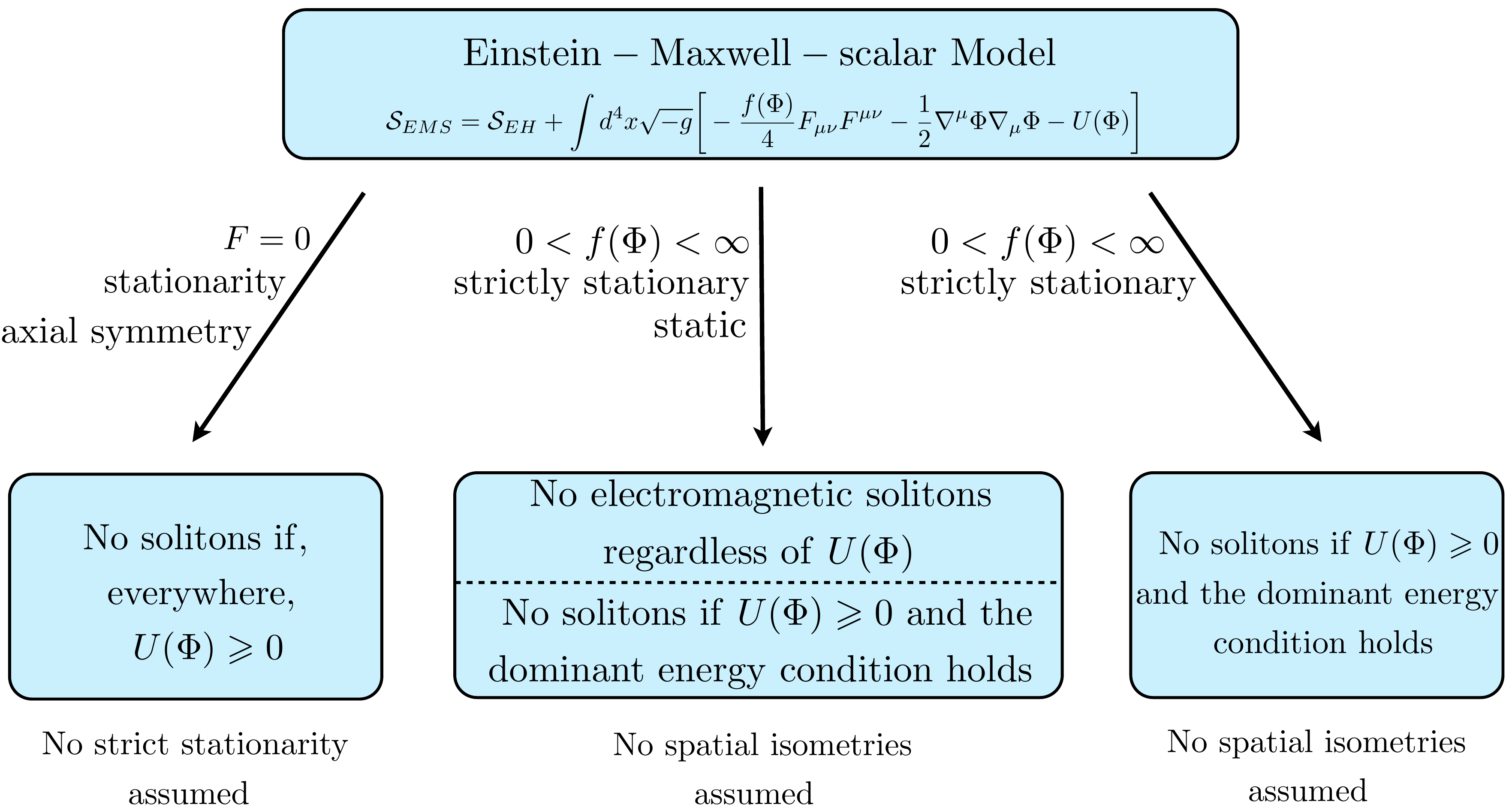}
	 	 \caption{Schematic representation summarising the no-scalar theorems presented herein. In all cases asymptotically flat spacetimes are assumed.}
	 	 \label{figure}
	\end{figure}

In this paper, motivated by the recently found ``hairy" black hole solutions in Einstein-Maxwell-scalar models~\cite{Herdeiro:2018wub}, which can arise dynamically from an instability of the RN black hole, we have addressed the existence of self-gravitating solitons in this family of models. We obtained three results for asymptotically flat spacetimes. The first is for the absence of axisymmetric and stationary solitons with a vanishing Maxwell field, as long as the scalar potential is everywhere non-negative and another for the absence of static (without spatial symmetries assumed) scalar-electromagnetic solitons in the full model. In this case no strict stationarity is assumed. The second theorem applies for strictly stationary and static spacetimes but without any assumptions on the spatial symmetries. If the coupling function $f(\Phi)$ does not change sign, then no electromagnetic solitons exist, regardless of scalar potential. Moreover, if one assumes the scalar potential to be non-negative and the dominant energy condition to hold, no scalar solitons exist either. Finally, the third result generalises the second by dropping the staticity assumption. A summary of these results is presented in a schematic way in Figure~\ref{figure}.

As a corollary of the results herein, the model studied in~\cite{Herdeiro:2018wub} illustrates that both bald and hairy black holes can exist without the existence of solitons. Therefore, even in models allowing both these sorts of black holes, not all hairy black holes can be faced as a superposition of a soliton and a bald black hole.

As a possible avenue for further work, it would be interesting to assess whether soliton solutions are actually possible if either $f(\Phi)$ changes sign or if $f(\Phi)$ diverges, say, at the origin. The physical interpretation of such models may, however, be a thorny issue.

\section*{Acknowledgements}
We would like to thank E. Ay\'on-Beato, F. Mena and, especially, E. Radu, for enlightening discussions and suggestions. J.O. is supported by the FCT grant PD/BD/128184/2016. 
This work has been supported by FCT (Portugal) through: the IF programme, grant PTDC/FIS-OUT/28407/2017, 
the strategic project UID/MAT/04106/2019 (CIDMA) and the CENTRA strategic project UID/FIS/00099/2013. 
We also acknowledge support from  the  European  Union's  Horizon  2020  
research  and  innovation  (RISE) programmes H2020-MSCA-RISE-2015
Grant No.~StronGrHEP-690904 and H2020-MSCA-RISE-2017 Grant No.~FunFiCO-777740. 
The authors would like to acknowledge
networking support by the
COST Action CA16104.

\end{document}